\documentclass[lettersize,journal]{IEEEtran}
\usepackage{amsmath,amsfonts}
\usepackage{algorithmic}
\usepackage{array}
\usepackage{textcomp}
\usepackage{booktabs}
\usepackage{tabularx}
\usepackage{pifont}
\usepackage{dblfloatfix}
\usepackage{multirow}
\usepackage{url}
\usepackage{float}
\usepackage{verbatim}
\usepackage{graphicx}
\usepackage{enumitem}
\usepackage{hyperref}

\ifCLASSOPTIONcompsoc
  \usepackage[caption=false,font=normalsize,labelfont=sf,textfont=sf]{subfig}
\else
  \usepackage[caption=false,font=footnotesize]{subfig}
\fi

\usepackage{xcolor} 
\usepackage[normalem]{ulem} 

\def\BibTeX{{\rm B\kern-.05em{\sc i\kern-.025em b}\kern-.08em
    T\kern-.1667em\lower.7ex\hbox{E}\kern-.125emX}}
\usepackage{balance}
\begin{document}
\title{RadarVox: Radar-Audio Multimodal Cocktail-Party Speech Separation with Speaker-Aware Cross-Modal Matching}
% \author{Yanlin~Xu,
% 		Yiwei~Ru,~\IEEEmembership{Member,~IEEE},
%         Mupei~Li,
%         Yongji~Liu,
%         Jie~Wang,~\IEEEmembership{Member,~IEEE},
%         and~Zhenan~Sun,~\IEEEmembership{Senior~Member,~IEEE}
% \thanks{Yanlin Xu is with the Institute of Automation, Chinese Academy of Sciences, Beijing, China (e-mail: yanlin.xu@cripac.ia.ac.cn).}
% \thanks{Yiwei Ru is with the Institute of Automation, Chinese Academy of Sciences, Beijing, China, and with the Beijing University of Posts and Telecommunications, Beijing, China (e-mail: yiwei.ru@cripac.ia.ac.cn).}
% \thanks{Yongji Liu is with the Beijing University of Posts and Telecommunications, Beijing, China (e-mail: yongjiliu@bupt.edu.cn; ). }
% \thanks{Jie Wang is with the School of Information Science and Technology, Dalian Maritime University, Dalian 116026, China (e-mail: wang\_jie@dlmu.edu.cn)}
% \thanks{Zhenan Sun and Mupei Li are with the Institute of Automation, Chinese Academy of Sciences, Beijing, China (e-mail: mupei.li@cripac.ia.ac.cn; znsun@nlpr.ia.ac.cn). \\Yiwei Ru is the corresponding author.}
% }
\author{Yanlin~Xu,
        Yiwei~Ru\textsuperscript{*},~\IEEEmembership{Member,~IEEE},
        Mupei~Li,
        Yongji~Liu,
        Jie~Wang,~\IEEEmembership{Member,~IEEE},
        and~Zhenan~Sun,~\IEEEmembership{Senior~Member,~IEEE}
\thanks{Yanlin Xu and Mupei Li are with the New Laboratory of Pattern Recognition (NLPR), State Key Laboratory of Multimodal Artificial Intelligence Systems (MAIS), Institute of Automation, Chinese Academy of Sciences (CASIA), Beijing 100190, China, and also with the School of Artificial Intelligence, University of Chinese Academy of Sciences (UCAS), Beijing 100049, China (e-mail: yanlin.xu@nlpr.ia.ac.cn ; mupei.li@cripac.ia.ac.cn).}
\thanks{Yiwei Ru is with the Institute of Automation, Chinese Academy of Sciences, Beijing 100190, China, and also with the Beijing University of Posts and Telecommunications, Beijing 100876, China (e-mail: yiwei.ru@cripac.ia.ac.cn).}
\thanks{Yongji Liu is with the Beijing University of Posts and Telecommunications, Beijing 100876, China (e-mail: yongjiliu@bupt.edu.cn).}
\thanks{Jie Wang is with the School of Information Science and Technology, Dalian Maritime University, Dalian 116026, China (e-mail: wang\_jie@dlmu.edu.cn).}
\thanks{Zhenan Sun is with the New Laboratory of Pattern Recognition (NLPR), State Key Laboratory of Multimodal Artificial Intelligence Systems (MAIS), Institute of Automation, Chinese Academy of Sciences (CASIA), Beijing 100190, China (e-mail: znsun@nlpr.ia.ac.cn).}
\thanks{\textsuperscript{*}Corresponding author: Yiwei Ru.}
}

% \markboth{Journal of \LaTeX\ Class Files,~Vol.~18, No.~9, September~2020}%
% {How to Use the IEEEtran \LaTeX \ Templates}

\maketitle

% \begin{abstract}
% This document describes the most common article elements and how to use the IEEEtran class with \LaTeX \ to produce files that are suitable for submission to the Institute of Electrical and Electronics Engineers (IEEE).  IEEEtran can produce conference, journal and technical note (correspondence) papers with a suitable choice of class options.
% \end{abstract}
\begin{abstract}

In embodied voice interaction, cocktail-party speech perception requires both speech separation and speaker attribution across machine-generated and human speech sources. However, conventional audio-only blind source separation remains permutation ambiguous, making the correspondence between separated streams and physical speakers unclear. This paper presents RadarVox, a radar-audio multimodal benchmark for identity-aware cocktail-party speech separation. RadarVox provides acoustic mixtures and source-level radar displacement signals from loudspeaker-emitted and human speech, enabling source-aware speaker assignment. FMCW radar captures laryngeal mechanical motion, providing speaker-specific spatial-motion cues unavailable to a single-channel microphone. We inject radar-derived priors into a DPRNN separator via gated fusion and learn a speaker-aware cross-modal matcher to associate unordered speech streams with radar-observed speakers. Experiments on multi-speaker mixtures show that radar cues provide complementary benefits, achieving scale-invariant signal-to-distortion ratio (SI-SDR) values of 9.75 dB and 7.13 dB in two- and three-speaker scenarios, respectively. More importantly, the proposed method improves ordered SI-SDR by more than 13 dB over audio-only methods while achieving over 80\% speaker assignment accuracy.

\end{abstract}


% \section{Introduction}
% \IEEEPARstart{A}{s} embodied intelligent systems evolve from single-agent task-execution platforms toward multi-agent collaborative systems deployed in real-world environments, their operational paradigm increasingly involves frequent interactions among agents as well as between agents and humans. In such interaction-intensive scenarios, speech serves as a primary communication medium for cooperative decision-making and information exchange. However, simultaneous vocalization from multiple speakers is nearly unavoidable in open environments. In the single-channel acoustic observation, speech signals from different sources are superimposed after propagation and further corrupted by environmental noise and reverberation due to the dependence of sound propagation on air medium. The microphone-recorded waveform then provides only a mixture observation, while the underlying contributions of multiple speakers remain latent. Recovering individual speech signals from a single-channel mixture therefore constitutes an underdetermined and non-unique inverse problem. As the number of concurrent speakers increases, overlap intensifies rapidly and separation ambiguity escalates, rendering the classical cocktail-party problem a fundamental challenge for embodied interactive systems operating in the real world as shown in Fig.~\ref{fig:intro}. Thus, the cocktail-party problem including determining "who said what" in a noisy environment with multiple speakers is the primary problem this research concentrates on.
\section{Introduction}
\begin{figure}[t]
    \centering
    \includegraphics[width=\linewidth,height=5.8cm]{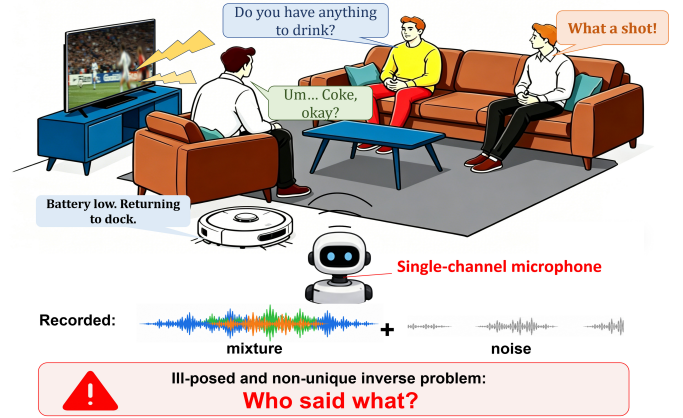}
    \caption{An example application scenario for speaker-aware speech perception. In daily environments with multiple speakers and ambient noise, a robot needs to perceive speaker-specific speech and provide personalized services accordingly.}
    \label{fig:intro}
\end{figure}

\IEEEPARstart{T}{he} cocktail-party problem is a long-standing challenge in multimedia speech perception, requiring systems to separate overlapping speech and associate each stream with its physical source \cite{11353938,2intro8656587,wham6813696}. This requirement is particularly relevant to embodied voice-interaction scenarios, such as homes with robots or smart devices, where machine-generated speech and human speech may coexist as illustrated in Fig.~\ref{fig:intro}.
In such shared acoustic spaces, simultaneous speech from multiple physical sources makes the problem more than signal separation alone: the system must determine not only what was said but who or what produced it. With a single-channel microphone, speech signals from different sources are superimposed and further degraded by environmental noise and reverberation \cite{ozturk2023radio}. The recorded waveform is therefore only a mixture observation, while source-specific contributions and their physical correspondence remain latent. Recovering source-specific speech from such a mixture is therefore an underdetermined inverse problem. As the number of concurrent sources increases, both acoustic overlap and output-source ambiguity become more severe, making identity-aware cocktail-party separation particularly challenging. Thus, an effective system should jointly address speech separation and physical source assignment.

% In light of the susceptibility of single-microphone recordings to noise contamination and their inability to resolve concurrent speakers, deep learning-based networks have been introduced to address these limitations, yet most of them target at blind source separation (BSS). Representative separation paradigms include deep clustering and permutation-invariant training (PIT), as well as end-to-end architectures base on PIT such as Conv-TasNet and DPRNN, which estimate source-specific signals directly from the acoustic mixture. However, even when multiple streams can be separated, these models do not natively determine which output corresponds to which physical speaker without auxiliary cues of sound sources. Meanwhile, this task is further challenged by the susceptibility of microphone recordings to environmental interference, together with the absence of explicit noise suppression in the training of these audio-only networks.

\begin{table*}[t]
\centering
\footnotesize 
\setlength{\tabcolsep}{3pt} 
\renewcommand{\arraystretch}{1.1}
\caption{Comparison of Existing Speech Datasets related to cocktail-party problem Across Modalities and Tasks}
\label{tab:dataset_comparison}
\resizebox{\textwidth}{!}{%
\begin{tabular}{l c >{\centering\arraybackslash}p{2.5cm} >{\centering\arraybackslash}p{1.5cm} >{\centering\arraybackslash}p{3.5cm} >{\centering\arraybackslash}p{2.5cm} >{\centering\arraybackslash}p{2.5cm} >{\centering\arraybackslash}p{1.2cm}}
\toprule
\textbf{Dataset} & \textbf{Sensing Modality} & \textbf{Task} & \textbf{Speech Source} & \textbf{Per-Speaker Text Coverage} & \textbf{Dataset Size} & \textbf{Corpus} & \textbf{Public Available} \\
\midrule
LibriMix \cite{cosentino2020librimix}      &Audio &Speech Separation&Human &Not applicable &364 h (921 speakers set) & LibriSpeech train-360&\ding{51} \\
%IEMOCAP \cite{busso2008iemocap}       &Audio-Visual &Emotion Recognition &Human & Selected emotional scripts + improvised hypothetical scenarios &$\sim$12 h (10 actors) &IEMOCAP &\ding{51} \\
Radio2Speech \cite{zhao2022radio2speech}  &Radar &Speech Recovery &Loudspeaker & LJSpeech corpus + TIMIT corpus&27.5 h (loudspeaker playback) & LJSpeech + TIMIT &\ding{55} \\
Wavesdropper \cite{wang2022wavesdropper}  &Radar& Closed-set Word Recognition&Human&57 fixed words &- (23 speakers)&57-word list &\ding{55} \\
WaveEar  \cite{xu2019waveear}     &Radar&Speech Recovery & Human&5 shared fixed passages ($\sim$1.5k words), 3 repetitions &$\sim$12 h (21 speakers) &5 reading materials & \ding{55}\\
Wavoice \cite{liu2024wavoice}      &Radar–Audio & Speech Recognition&Human &40 shared commands,40 repetitions&- (20 speakers) &ok-google.io + Google speech commands & \ding{55}\\
mmMUSE \cite{wang2025mmmuse}       &Radar–Audio & Speech Enhancement&Human & 15 commands + 13 sentences,10 repetitions& 15 h (46 speakers)&ok-google.io + TIMIT subsets & \ding{55} \\
RadioSES \cite{ozturk2023radio}     &Radar–Audio&Speech Enhancement, Speech Separation &Human &300 sentences&- (19 speakers) &TIMIT &\ding{55} \\
\midrule
RadarVox          &Radar–Audio &Speech Recovery, Speech Enhancement, Identity-aware Speech Separation, Emotion Recognition &Human, Loudspeaker & 300 sentences, 210 sentences per loudspeaker timbre&16 h (before generation, 19 speakers + loudspeaker playback) &TIMIT + self-designed corpus for the loudspeaker &\ding{51} \\
\bottomrule
\end{tabular}%
}
\end{table*}

Existing studies have made substantial progress on the cocktail-party problem through audio-only blind source separation (BSS). Representative paradigms include deep clustering, permutation-invariant training (PIT), and end-to-end architectures such as Conv-TasNet and DPRNN, which estimate source-specific waveforms directly from acoustic mixtures \cite{8707065,luo2020dual,hershey2016deep,yu2017permutation}. These methods have established strong baselines for single-channel speech separation. However, their outputs are inherently unordered, and they do not explicitly determine which separated stream corresponds to which physical source.

% Given the limited source information available from single microphone recordings and their susceptibility to noise interference, together with the inability of purely acoustic networks to achieve source aware separation, inspiration is drawn from the human auditory solution to the cocktail party problem. Despite the inevitable acoustic overlap in multi-talker environments, human auditory system remains remarkably robust. Psychoacoustic and neurophysiological studies suggest that speech segregation in cocktail-party settings is not only achieved by monaural acoustic analysis on cues related to pitch, harmonicity, timbre, and temporal continuity. Instead, the auditory system exploits a combination of binaural spatial cues, such as interaural time differences (ITD) and interaural level differences (ILD). By contrast, single microphone relies only on a monaural acoustic mixture and lack direct access to auxiliary physical observations informative of source location or speaker-specific production cues. 

To transcend the inherent limitations of purely acoustic BSS networks, we draw inspiration from the human auditory system. Human cocktail-party perception relies heavily on binaural spatial cues, such as interaural time differences (ITD) and interaural level differences (ILD), to segregate competing sound sources \cite{franken2015vivo}. Since such binaural cues are unavailable to a single-channel microphone, frequency-modulated continuous-wave (FMCW) radar is introduced to provide range-resolved spatial-motion constraints for separation and assignment. At the source level, reflections from different physical sources can be mapped to distinct range bins, offering spatial cues for output-source correspondence. At the motion level, radar-derived displacement captures phonation-induced mechanical vibration, such as human laryngeal surface micro-motion or loudspeaker diaphragm motion, which is coupled with acoustic speech generation but sensed through electromagnetic reflection rather than air-borne sound propagation. This propagation difference makes radar less sensitive to environmental acoustic noise and complementary to microphone recordings. Therefore, radar can provide spatial-motion cues for improving separation under overlap and noise, while also supporting explicit output-source correspondence.

However, radar-audio identity-aware separation remains limited by the lack of open-access datasets that provide acoustic mixtures, source-level radar observations, and explicit source correspondence. As summarized in Table~\ref{tab:dataset_comparison}, audio-only datasets such as LibriMix lack auxiliary physical observations %while audio-visual datasets introduce useful non-acoustic cues but may be constrained by privacy concerns and illumination conditions 
\cite{cosentino2020librimix}. Existing radar-based speech datasets are often limited by heterogeneous radar configurations, restricted speech corpora such as command words or short phrases, and single-source recording protocols, making them insufficient for multi-source separation with explicit radar correspondence. Critically, no publicly available radar-audio dataset has been explicitly designed for identity-aware cocktail-party speech separation, where separated streams must be associated with their corresponding physical sources.

To address these limitations, we construct RadarVox, a radar-audio multimodal benchmark tailored for identity-aware cocktail-party speech separation. RadarVox dataset covers two complementary speech-source settings: loudspeaker-emitted speech and natural human speech at \href{https://doi.org/10.21227/6sw9-5p65}{DOI:10.21227/6sw9-5p65}. The loudspeaker subset models machine-generated speech from devices or agents and provides a controllable source with stable radar-observable diaphragm vibration, enabling systematic radar parameter selection and preprocessing pipeline validation before human speech acquisition. The human subset records TIMIT utterances from 19 speakers, from which clean and noisy two- and three-speaker mixtures are generated for separation and source assignment \cite{2timit}. RadarVox dataset provides temporally aligned acoustic mixtures and source-level radar displacement sequences derived from speech-induced mechanical motion. Based on this dataset, we develop a two-stage framework that injects radar-derived motion priors into a DPRNN separator through gated fusion and learns a speaker-aware cross-modal matcher to resolve output-source ambiguity. To the best of our knowledge, this is the first radar-audio framework that explicitly learns source-level correspondence for identity-aware cocktail-party speech separation.

The main contributions of this work are summarized as follows.
\begin{enumerate}[label=\arabic*), topsep=0pt, itemsep=0pt, parsep=0pt, partopsep=0pt]
 \item We establish RadarVox, a comprehensive radar-audio dataset designed for identity-aware cocktail-party speech perception, featuring explicit source-level correspondence for diverse human-agent interactive scenarios.

 \item We formulate a tailored radar signal processing pipeline that maps raw FMCW measurements into fine-grained laryngeal and diaphragm displacement sequences via phase-aligned coherent accumulation.

 \item We propose a two-stage cross-modal framework unifying radar-guided speech enhancement and speaker-aware identity matching, which substantially mitigates the permutation ambiguity of purely acoustic separation and facilitates explicit output-source assignment.

\end{enumerate}

The remainder of this paper is organized as follows. Section II reviews related work. Section III introduces the RadarVox benchmark, including dataset construction, task protocols, and radar displacement extraction. Section IV presents the proposed two-stage radar-audio framework. Section V reports experimental results and analyses. Section VI concludes the paper.

\section{Related Work}
\subsection{Audio-Only Speech Separation}
Single-channel speech separation methods for the cocktail-party problem can be broadly categorized into time-frequency mask estimation and end-to-end time-domain modeling. Representative approaches include deep clustering and permutation-invariant training (PIT), which respectively learn source-discriminative embeddings and address label ambiguity by optimizing over possible output permutations \cite{hershey2016deep,yu2017permutation}. End-to-end time-domain methods further learn separation directly from waveforms with task-adaptive encoder. TasNet and Conv-TasNet establish encoder-separator-decoder frameworks for waveform-level separation \cite{8707065,luo2018tasnet}. DPRNN further improves long-sequence modeling through intra- and inter-chunk recurrent processing \cite{luo2020dual}.

These single-modal audio methods are designed for blind source separation (BSS) and rely on the statistical distinguishability of different sources within the acoustic mixture. In real-world cocktail-party scenarios, however, heavy spectral overlap, similar speaker characteristics, environmental noise, and reverberation can substantially weaken such distinguishability \cite{wham6813696,3wham9863655}.

\subsection{Radar-Only Speech Sensing}
The main difficulty in radar-only speech sensing is that micro-vibrations related to speech are extremely weak, rendering the recovery quality highly sensitive to the preprocessing pipeline. Existing radar-only studies mainly focus on two types of vibration sources: loudspeaker diaphragm motion and laryngeal surface micro-motion. Unlike microphones that record air-borne acoustic pressure and are directly affected by acoustic interference, radar senses speech-induced mechanical vibration through electromagnetic reflection and is less sensitive to environmental noise \cite{liu2024wavoice}. This property has motivated radar-only speech recovery, where speech-related motion is measured without relying on acoustic wave propagation. 

Early work explored speech recovery from loudspeaker diaphragm vibration, where the vibration source is relatively strong and stable. For example, RadioMic localizes loudspeaker vibration using range-Doppler representations and recovers speech-related signals through phase demodulation \cite{2radiomic}. A U-Net-based reconstruction module is further used to compensate for the limited bandwidth of radar-derived vibration measurements \cite{2radiomic}.

\begin{figure*}[t]
    \centering
    \includegraphics[width=\textwidth]{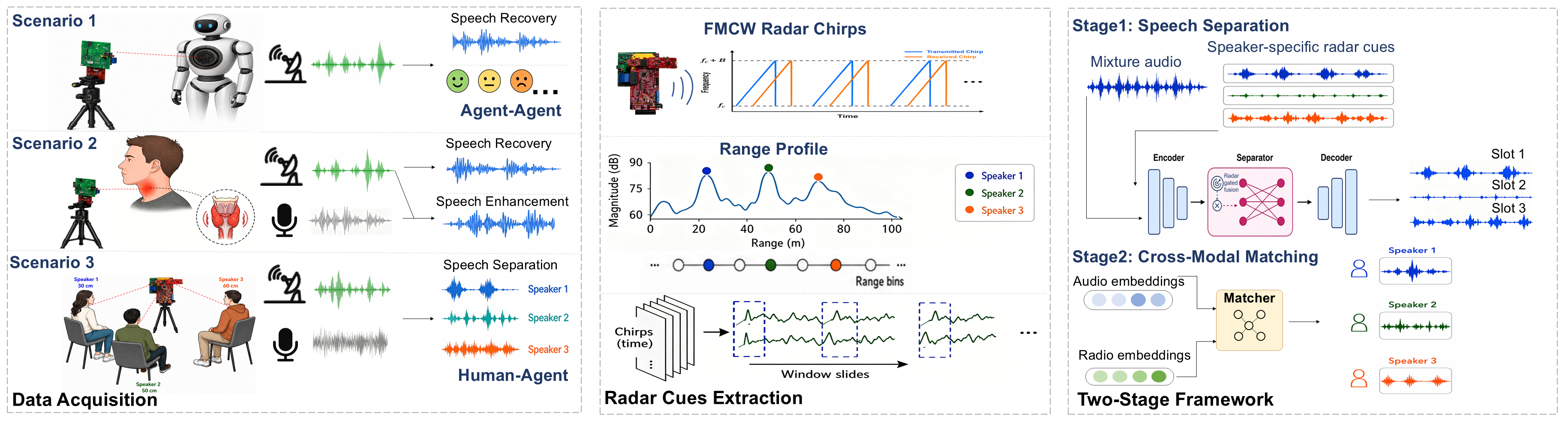}\par
    \makebox[\textwidth][c]{%
      \makebox[0.33\textwidth][c]{\hspace*{0.03\textwidth}(a)}%
      \makebox[0.33\textwidth][c]{\hspace*{0.04\textwidth}(b)}%
      \makebox[0.33\textwidth][c]{\hspace*{0.01\textwidth}(c)}%
    }
    \caption{Overview of the Proposed Radar--Audio Speech Perception System. (a) Radar acquisition of loudspeaker vibrations to simulate agent-agent communication, and acquisition of human laryngeal vibrations to simulate agent-human interaction. (b) The transmitted and received radar chirps are preprocessed to localize the speaker and extract the corresponding laryngeal micro-motion signal. (c) The proposed two-stage model for identity-aware speech separation.}
    \label{fig:system}
\end{figure*}

Compared with loudspeaker diaphragm motion, throat vibrations exhibit minute amplitudes and is more vulnerable to head motion and non-speech articulatory interference. This imposes stricter requirements on spatial localization and phase stability. WaveEar mitigates these challenges by beamforming-based throat localization, followed by an STFT-based residual convolutional network that maps mmWave spectrograms to speech spectrograms \cite{xu2019waveear}.

Despite these advances, standalone radar-only speech recovery remains limited because radar captures speech-induced mechanical motion rather than the radiated acoustic waveform. Loudspeaker diaphragm vibrations mainly reflect device-specific diaphragm motion, while human laryngeal vibrations are observed before vocal-tract modulation and thus lack part of articulatory details. Together with sampling-rate and bandwidth limits in practical mmWave systems, these factors make radar-only sensing insufficient for high-fidelity speech reconstruction.

Therefore, radar-only sensing is valuable for extracting source-level speech-related motion cues, but it is better suited as an auxiliary modality than as a standalone replacement for audio in cocktail-party separation.

\subsection{Radar-Audio Speech Enhancement and Separation}
Radar-audio fusion has first been explored mainly for single-speaker speech enhancement, where radar-derived motion cues complement noisy acoustic recordings. For example, mmMUSE combines motion compensation with a two-stage complex-valued cross-modal enhancement framework and formulates enhancement as complex ratio mask estimation in the time-frequency domain \cite{wang2025mmmuse}. By contrast, radar-audio studies on multi-speaker cocktail-party separation remain limited. RadioSES extends radar-audio fusion to multi-speaker separation using an end-to-end encoder-masker-decoder architecture with DPRNN-based temporal modeling \cite{ozturk2023radio}. Nevertheless, it still follows BSS formulation and does not associate separated outputs with radar-observed physical sources.

Overall, radar-audio fusion remains primarily focused on single-speaker enhancement, whereas cocktail-party separation with explicit output-source correspondence is still unexplored.

\section{RadarVox Benchmark for Identity-Aware Speech Separation}
%Fig.~\ref{fig:system} provides an overview of the proposed benchmark-to-method pipeline.
RadarVox is designed as a radar-audio benchmark for identity-aware cocktail-party separation, where separated streams should be associated with their corresponding physical sources. In addition to the main multi-speaker separation task, RadarVox supports single-speaker enhancement and extended modality analysis to characterize radar-derived speech-related motion cues. This section describes the recording scenarios, radar acquisition and preprocessing pipeline, and dataset composition.

%RadarVox is constructed as a radar-audio benchmark centered on identity-aware multi-speaker speech separation. While the released dataset also support speech recovery, speech enhancement, speech separation and emotion recognition, these tasks are included mainly to broaden benchmark coverage and to characterize the sensing modalities under different conditions. This section focuses on recording scenarios, radar acquisition configuration, preprocessing protocol, and dataset composition.

\subsection{Speech-Source Settings}

\begin{figure}[htbp]
    \centering
    \subfloat[]{
        \includegraphics[width=0.45\columnwidth,height=2.8cm]{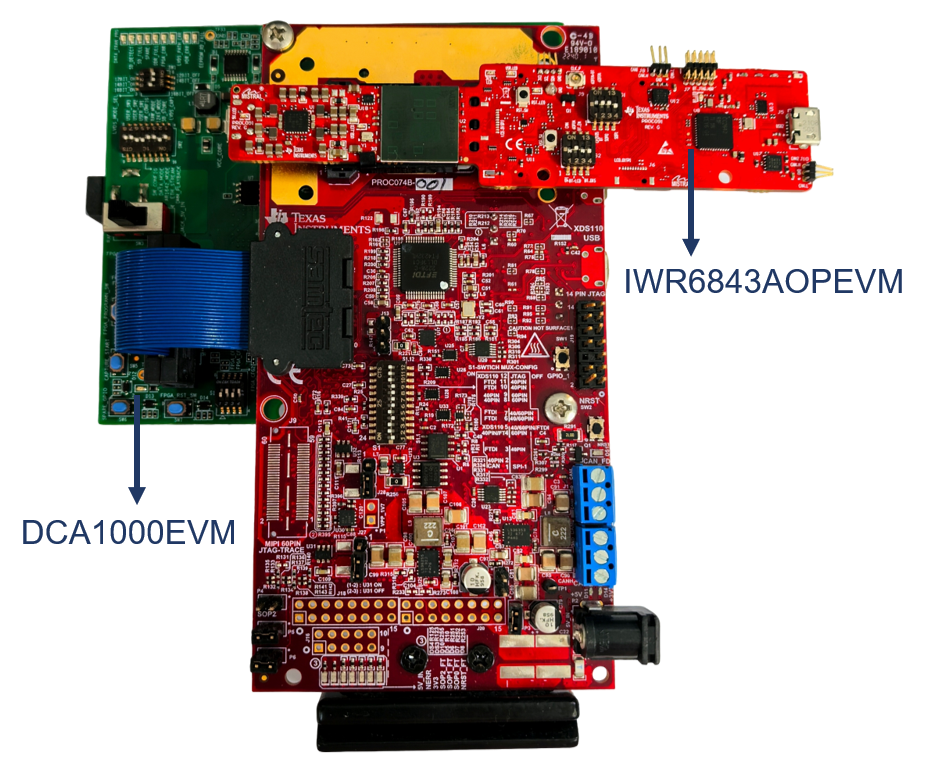}
        \label{fig:device_a}
    }
    \hfill
    \subfloat[]{
        \includegraphics[width=0.45\columnwidth,height=3cm]{device.pdf}
        \label{fig:device_robot}
    }
    \caption{Recording devices and scenarios of the loudspeaker. (a) Radar devices. (b) Recording scenarios of loudspeaker.}
    \label{fig:device_group_robot}
\end{figure}
\subsubsection{Loudspeaker-Emitted Speech}
A loudspeaker-based setting is introduced as a controlled condition for machine-generated speech. It provides a reproducible testbed for radar parameter selection and displacement-extraction validation before human speech recording. Compared with laryngeal surface micro-motion, loudspeaker diaphragm vibration is stronger and mechanically more stable, making it suitable for controlled validation of the radar sensing pipeline.
The loudspeaker and radar are mechanically isolated during acquisition to reduce platform vibration. as in Fig.~\ref{fig:device_robot}. The IWR6843AOPEVM module is oriented with its antenna-on-package (AoP) side facing the loudspeaker diaphragm. 

\subsubsection{Single-Speaker Human Speech}
For human speech sensing, the radar localizes the laryngeal region and captures subtle surface micro-vibrations induced by vocal-fold motion.
Unlike loudspeaker diaphragm vibration, laryngeal surface motion is weaker and less mechanically constrained. It reflects phonation-related surface motion before full vocal-tract filtering and articulation are expressed in the radiated acoustic waveform. Therefore, radar is used as an auxiliary physical modality, while the microphone provides the primary acoustic observation.

\begin{figure}[htbp]
    \centering
    \subfloat[]{
        \includegraphics[width=0.45\columnwidth,height=2.8cm]{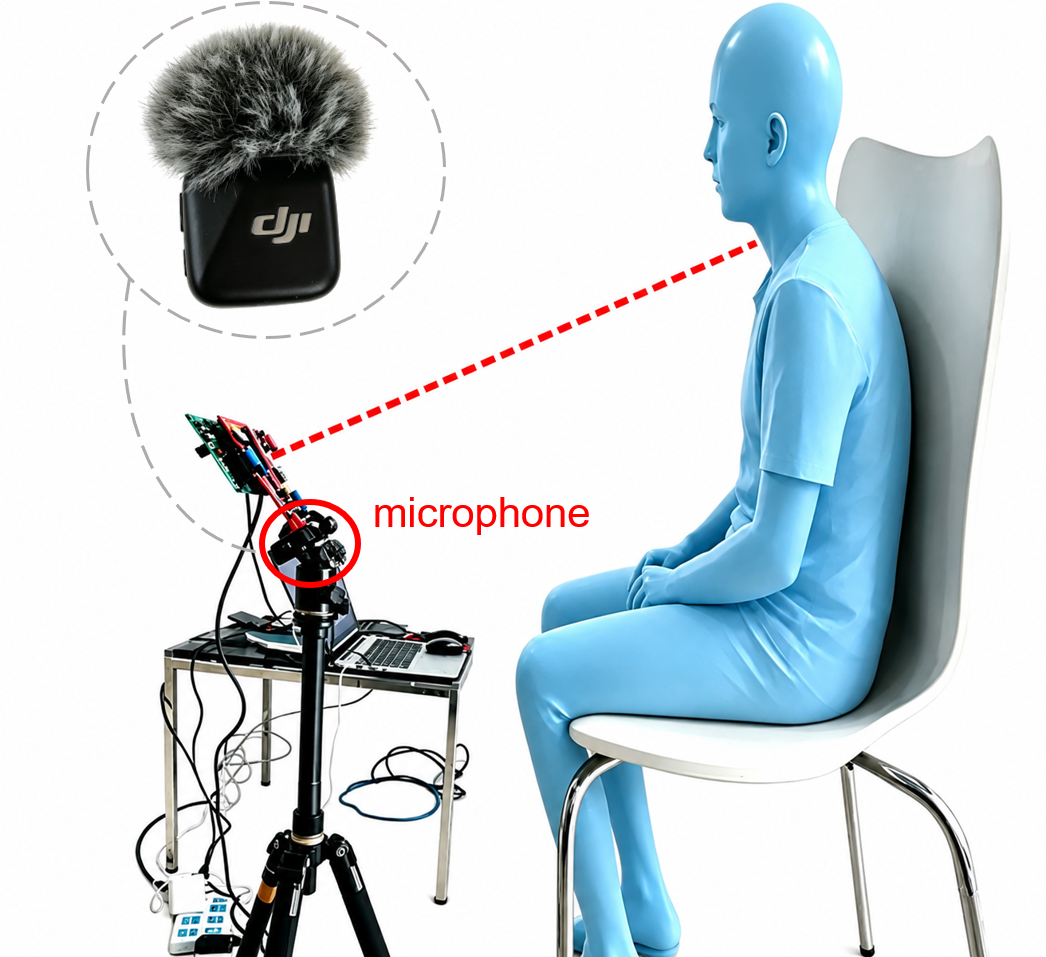}
        \label{fig:human001}
    }
    \hfill
    \subfloat[]{
        \includegraphics[width=0.45\columnwidth,height=2.8cm]{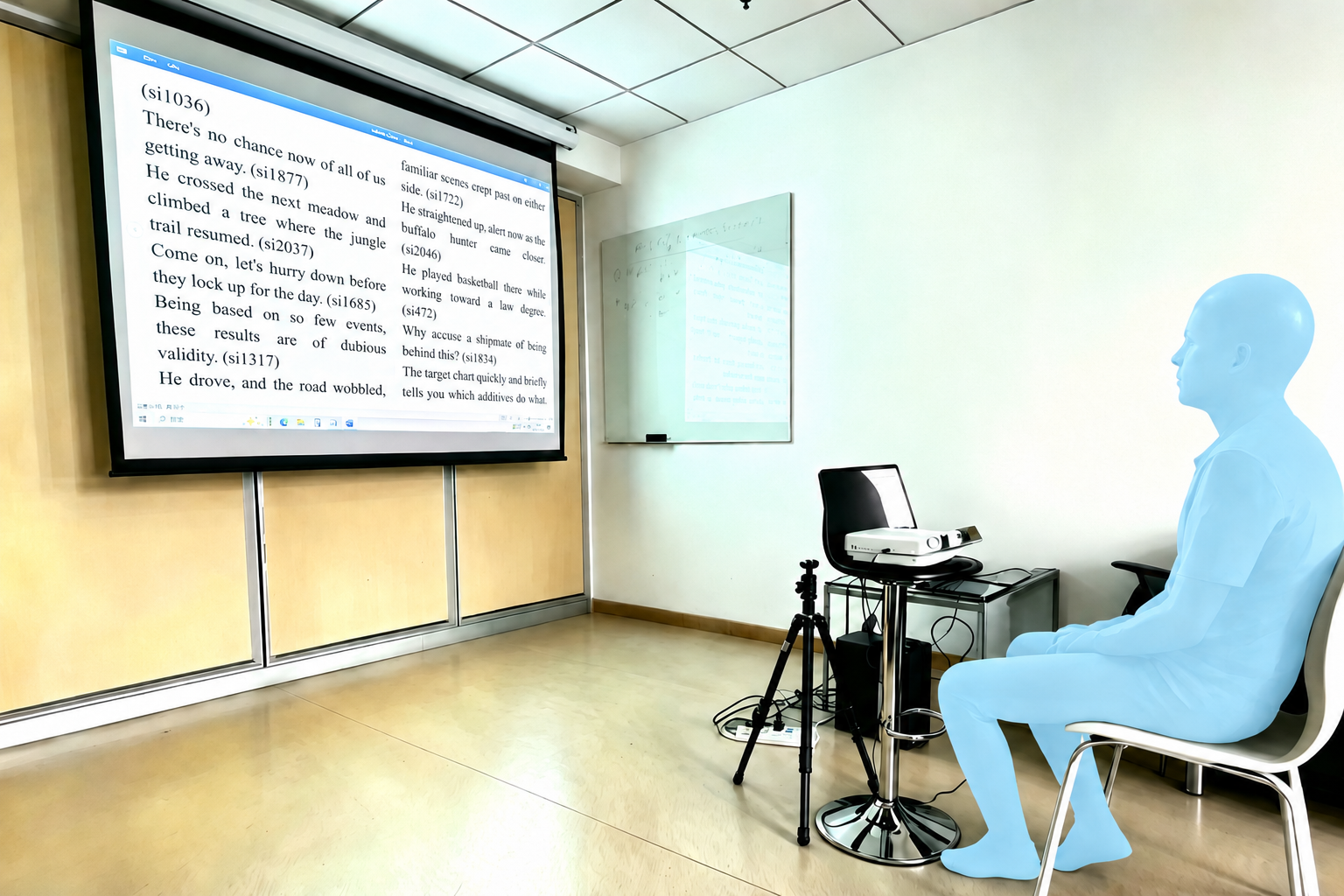}
        \label{fig:human002}
    }
    \caption{Recording scenarios of the speaker. (a) Orientation of radar. (b) Recording scenarios of the speaker.}
    \label{fig:device_group_human}
\end{figure}

In the single-speaker setting, the radar and microphone are oriented toward the speaker’s laryngeal region as shown in Fig.~\ref{fig:human001}.
Speakers read TIMIT sentences under synchronized radar and audio recording
as shown in Fig.~\ref{fig:human002} \cite{2timit}. Natural pauses, minor reading variations, and prosodic diversity are preserved to maintain realistic speech characteristics.

\subsubsection{Multi-Speaker Human Speech}
The main separation benchmark is constructed from single-speaker human recordings, from which two- and three-speaker mixtures are generated following the LibriMix protocol \cite{cosentino2020librimix}.Additional real overlapped recordings are collected with speakers seated at different distances from the radar to evaluate practical feasibility as shown in Fig.~\ref{fig:system}. 
%During data collection, the speakers speak with temporal overlap.

\subsection{Radar Acquisition Configuration}
Because loudspeaker diaphragm vibration is stronger and more repeatable than laryngeal surface motion, the controlled loudspeaker setting is used to validate radar parameters and preprocessing strategies.
\subsubsection{Parameter Design}
Temporal sampling rate is critical to radar-based speech acquisition because it determines the recoverable vibration bandwidth. Audio is recorded at 48 kHz using a DJI Mic Mini microphone. Most prior radar speech-sensing systems use slow-time sampling rates of 1--2 kHz, limiting the recoverable vibration bandwidth \cite{ozturk2023radio,2radiomic}. However, the articulation index, which quantifies speech intelligibility, falls below 50\% when speech is band-limited to 2 kHz \cite{2radiomic}.

To alleviate this limitation, instead of retaining only one chirp per frame as in conventional processing, the frame periodicity is increased and all in-frame chirps are preserved. Specifically, the received frame--chirp cube are unfolded into a chirp sequence as defined in Eq.~\eqref{eq:unfold}. The resulting configuration, summarized in Table~\ref{tab:radar_params}, yields an effective chirp-time sampling rate of approximately 11 kHz.

\begin{table}[t]
\centering
\caption{Radar System Parameter Configuration}
\label{tab:radar_params}
\begin{tabular}{l c}
\toprule
Parameter & Value \\
\midrule
Ramp End Time & 83~$\mu$s \\
Idle Time & 5~$\mu$s \\
Frequency Slope & 42~MHz/$\mu$s \\
Loop Count (No. of Chirps per Frame) & 255 \\
Frame Periodicity & 23~ms \\
\bottomrule
\end{tabular}
\end{table}

\subsubsection{Preprocessing Choices}
\begin{figure*}[t]
    \centering
    \includegraphics[width=\textwidth]{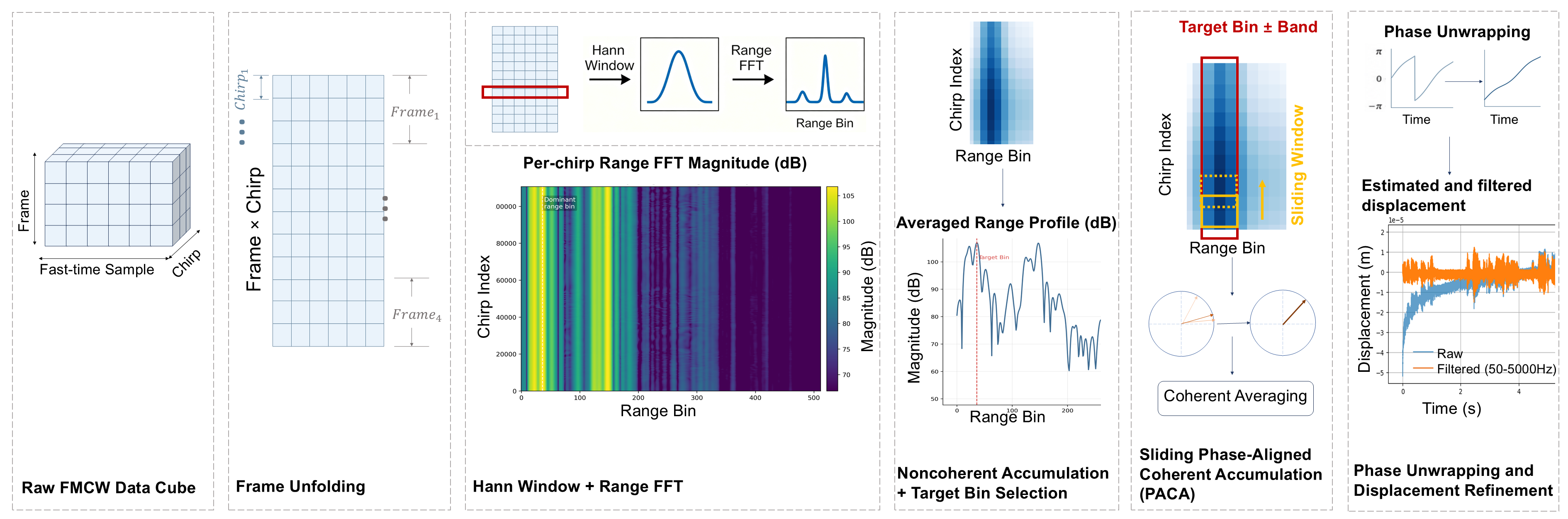}
    \caption{Schematic of the Radar Preprocessing Pipeline for Loudspeaker Data. The six sequential steps are illustrated from left to right.}
    \label{fig:pre}
\end{figure*}

As illustrated in Fig.~\ref{fig:pre}, the proposed preprocessing pipeline converts raw FMCW measurements into one-dimensional displacement sequences associated with speech-induced micro-vibration. Each chirp is transformed into a range profile after Hann windowing and fast-time Fourier transform (FFT) as in Eq.~\eqref{eq:fft}. Noncoherent accumulation across chirps is executed to identify the dominant speech-related scatterer as in Eq.~\eqref{eq:peak},  

\begin{equation}
x_k(n)=x(f,c,n),\qquad k=fC+c,
\label{eq:unfold}
\end{equation}
where \(x(f,c,n)\in\mathbb{C}\) denotes the raw mmWave signal indexed by frame \(f\), chirp \(c\), and fast-time sample \(n\), where \(F\) is the total number of frames, \(C\) is the number of chirps per frame, and \(S\) is the number of fast-time samples in each chirp. The indices satisfy \(f\in\{0,\dots,F-1\}\), \(c\in\{0,\dots,C-1\}\), and \(n\in\{0,\dots,S-1\}\), and \(k\) denotes the unfolded chirp index.

\begin{equation}
X_k(r)=\sum_{n=0}^{S-1}x_k(n)w(n)e^{-j2\pi rn/N},
\label{eq:fft}
\end{equation}
where \(w(n)\) is the Hann window, \(N\) is the FFT size, and \(r\) denotes the range-bin index.

\begin{equation}
r^{*}=\arg\max_{r\in[0,N/2]}\frac{1}{K}\sum_{k=0}^{K-1}\lvert X_k(r)\rvert,
\label{eq:peak}
\end{equation}
where \(r^{*}\) is the selected target range bin obtained by non-coherent accumulation and \(K\) is the total number of unfolded chirps.

Subsequently, the core operation is phase-aligned coherent accumulation (PACA). PACA performs local phase alignment and coherent averaging within sliding windows, improving temporal consistency without assuming global phase stationarity. Specifically, for the \(m\)-th sliding window \(W_m=\{mH,\dots,mH+L-1\}\), where \(L\) and \(H\) denote the window length and hop size, respectively, the coherently integrated spectrum is given by 
\begin{equation}
Y_m(r) = \frac{1}{L} \sum_{k \in W_m} X_k(r) \exp\!\left(-j\bigl(\angle X_k(r^*) - \phi_m^{\mathrm{ref}}\bigr)\right),
\end{equation}
where $\phi_m^{\mathrm{ref}} = \operatorname{median}\{\angle X_k(r^*) \mid k \in W_m\}$.

%This formulation assumes local phase consistency of the speech-induced micro-vibration within each window, enabling constructive accumulation of the vibration component while mitigating phase incoherence across chirps.

The aligned slow-time phase is then unwrapped and converted into radial displacement via
\begin{equation}
d(t)=\frac{\lambda}{4\pi}\phi(t),
\label{eq:displacement}
\end{equation}
where $\lambda$ denotes the radar wavelength. 

% The resulting displacement signal is then bandpass filtered from 50~Hz to 5000~Hz to isolate physiologically relevant speech-induced vibration components and resampled to 16~kHz to obtain a baseband signal for downstream neural speech enhancement.

\begin{table}[t]
\centering
\caption{
Speech reconstruction performance for a loudspeaker under different radar preprocessing strategies.
}
\label{tab:slowtime_preproc_compact}
\setlength{\tabcolsep}{6pt}
\renewcommand{\arraystretch}{1.15}
\begin{tabular}{l c c c }
\toprule
\textbf{Preprocessing Strategy} 
& \textbf{SI-SDR} $\uparrow$ 
& \textbf{PESQ} $\uparrow$ 
& \textbf{STOI} $\uparrow$ \\
\midrule
EMA Clutter Cancellation  \cite{ali2021goodness}                         & -38.10 & 1.027 & 0.177 \\
VMD-based Denoising \cite{fang2023denoising}                               & -56.37 & 1.367 & 0.178 \\
Wavelet Shrinkage Denoising \cite{fang2023denoising}                       & -42.76 & 1.033 & 0.066 \\
Kalman Smoothing  \cite{20kal8656587}                                 & -41.06 & 1.020 & 0.158 \\
\midrule
Phase-aligned Coherent Accumulation                 & \textbf{-37.95} & 1.040 & 0.104 \\
\bottomrule
\end{tabular}
\end{table}

We compare PACA with exponential moving average (EMA) clutter cancellation, variational mode decomposition (VMD), wavelet shrinkage denoising, Kalman smoothing \cite{ali2021goodness,fang2023denoising,20kal8656587,lies2021longPACA}. Signal-to-distortion ratio (SI-SDR), perceptual evaluation of speech quality (PESQ), and short-time objective intelligibility (STOI) are used as relative downstream proxy metrics. As shown in Table~\ref{tab:slowtime_preproc_compact}, PACA achieves the best SI-SDR, suggesting better preservation of speech-related temporal structure.

\subsection{Radar-Sensing Capabilities for Speech Separation}
This section highlights two radar-sensing capabilities of RadarVox dataset for speech separation: high effective chirp-time sampling rate and range-based source separability.
\subsubsection{Chirp-Time Sampling rate}
RadarVox provides a high effective chirp-time sampling rate, which improves the temporal resolution of radar-derived motion signals. With PACA, the effective sampling rate is determined by the number of chirps per frame, the frame rate \(f_{\mathrm{frame}}\), and the hop size \(H\):
\begin{equation}
f_s = \frac{C \cdot f_{\mathrm{frame}}}{H}.
\end{equation}

As shown in Table~\ref{tab:sampling_rate}, higher sampling rate substantially improves radar-derived speech representations, with the 11087 Hz configuration yielding markedly lower WER and stronger SI-SDR than lower-rate settings. These results confirm the importance of dense chirp-time sampling for preserving speech-related motion cues.

\begin{table}[t]
\centering
\caption{Performance Under Different Sampling Rates}
\label{tab:sampling_rate}
\resizebox{\columnwidth}{!}{
\begin{tabular}{c c c c c c}
\toprule
Sampling Rate (Hz) & SI-SDR (dB) & STOI& PESQ &  SNR (dB) & WER\\
\midrule
1000  & -51.29 & \textbf{0.169} &1.036 &  -3.02 & 1.00\\
7000  & \textbf{-32.63} & 0.048 &\textbf{1.033} &  \textbf{-2.76} & 1.00\\
11087 & -37.95 & 0.104 &1.040 &  -2.82 & \textbf{0.41}\\
\bottomrule
\end{tabular}
}
\end{table}

\subsubsection{Range-Based Source Separability}
A single FMCW radar can map reflections from different source distances to distinct range bins, providing range-separated motion signatures associated with speech-induced micro-vibrations. Such range-based separability provides geometric constraints for associating separated streams with their corresponding physical sources. With the configuration in Table~\ref{tab:radar_params}, RadarVox achieves a range resolution of 4.30 cm.

\subsection{Dataset Corpus and Split Protocol}
The first part of RadarVox dataset is designed to simulate embodied agent--agent interactions through radar sensing of loudspeaker-emitted speech. A dedicated loudspeaker-based corpus is constructed by generating sentences longer than 25 characters with DeepSeek under constraints of syntactic completeness, semantic richness, and everyday relevance \cite{2guo2025deepseek}. Seven emotion categories are included for extended modality analysis. The text is converted into emotion-conditioned speech using Index-TTS2. \cite{zhou2026indextts2}. Ten reference voices are selected to ensure timbre diversity. The corpus is partitioned into training, validation, and test sets using mutually disjoint speaker identities, with 7, 1, and 2 timbres assigned to the three splits, respectively.

The second part is designed to simulate human--agent interactions and uses TIMIT as the speech corpus \cite{2timit}. Utterances shorter than 25 characters are filtered to retain sufficient temporal and phonetic coverage. Each speaker contributes approximately 300 utterances. The corpus includes 19 speakers and is divided into speaker-disjoint training, validation, and test splits with 11, 4, and 4 speakers, respectively. The cohort comprises 12 male and 7 female speakers, with natural variation in accent and pausing patterns. Informed consent is obtained from all volunteers and the collected data are anonymized before research use.
%The selected material includes \textit{SA} sentences for cross-speaker normalization, \textit{SX} sentences with compact phonetic coverage, and \textit{SI} sentences with speaker-specific phonetic variability \cite{garofolo1993timit}. Each speaker reads two shared \textit{SA} sentences together with randomly selected \textit{SX} and \textit{SI} sentences, yielding approximately 300 utterances per speaker. In total, 19 speakers are included, and the dataset is divided into mutually disjoint speaker-level training, validation, and test splits containing 11, 4, and 4 speakers, respectively. The cohort comprises 12 male and 7 female speakers, with natural variation in accent and pausing patterns. Informed consent is obtained from all volunteers and the collected data are anonymized before research use.

\begin{figure}[t]
    \centering
    \includegraphics[width=\linewidth,height=3.9cm]{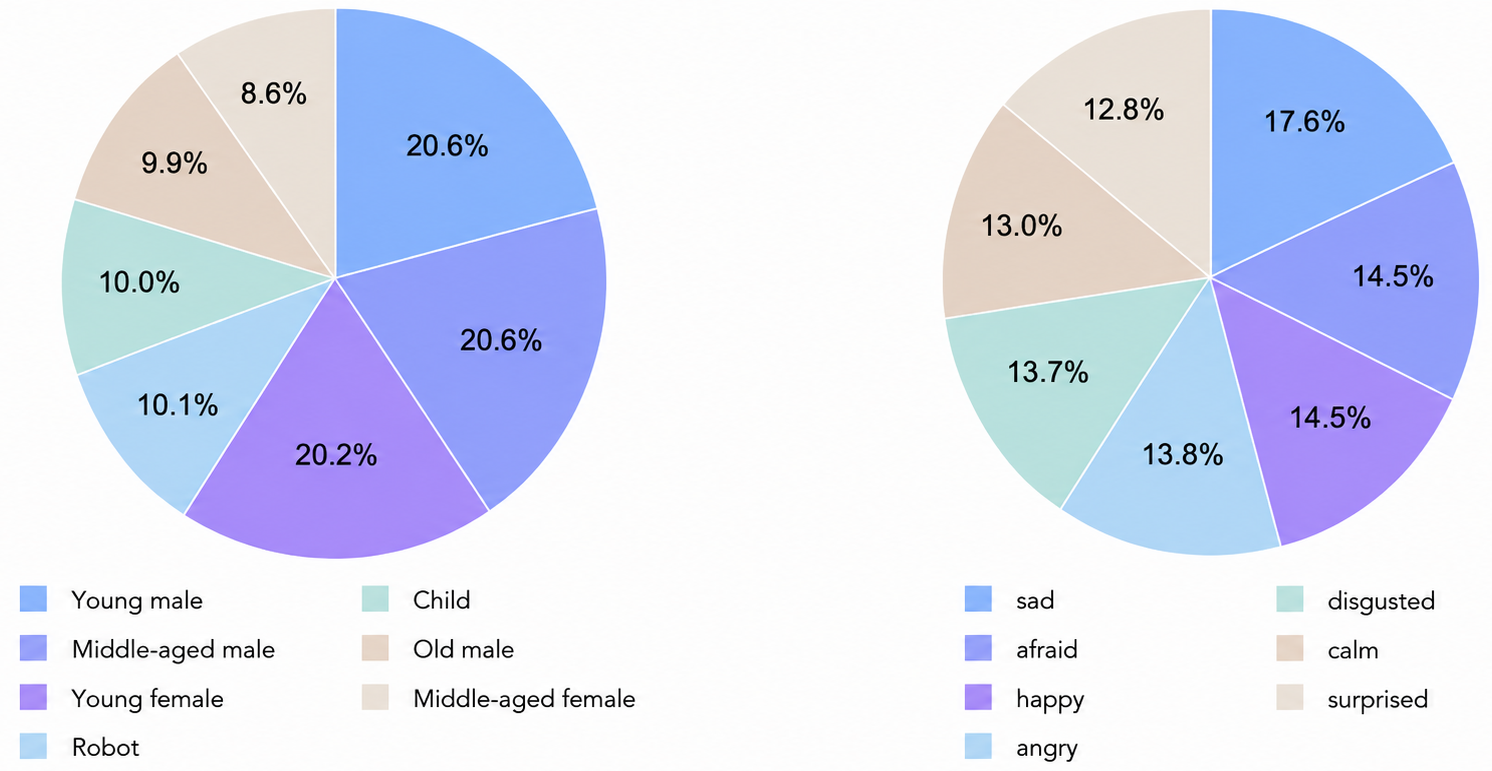}\par
    {\footnotesize
    \makebox[0.5\linewidth][c]{(a)}%
    \makebox[0.5\linewidth][c]{(b)}
    }
    \caption{Distributions of loudspeaker's recordings across voice categories and emotions, measured by total duration. (a) Timbre-category distribution. (b) Emotion distribution.}
    \label{fig:pies_robot}
\end{figure}

\begin{figure}[t]
    \centering
    \subfloat[]{%
        \includegraphics[width=0.48\columnwidth]{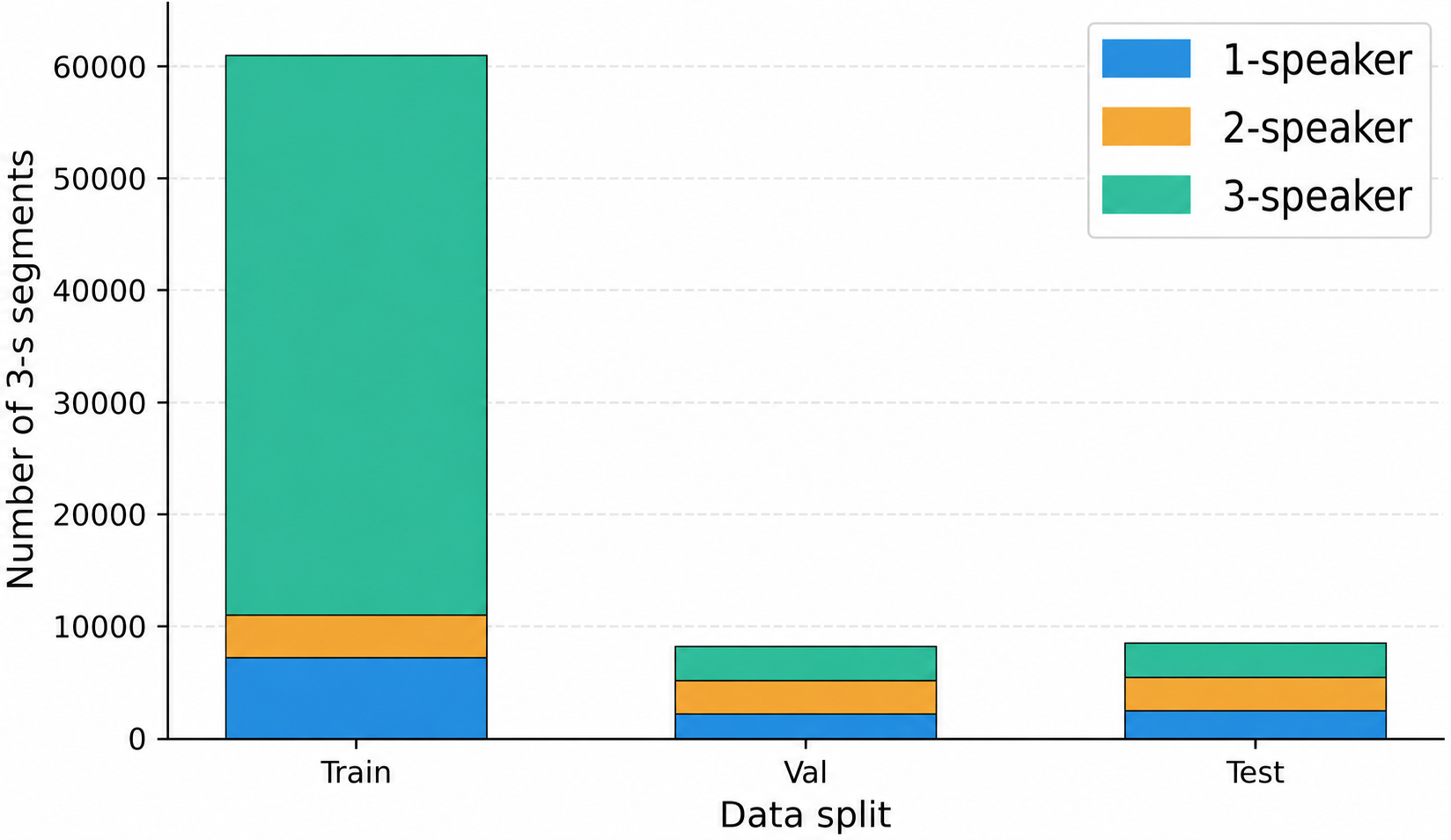}
        \label{fig:data_bar}
    }
    \hfill
    \subfloat[]{%
        \includegraphics[width=0.48\columnwidth]{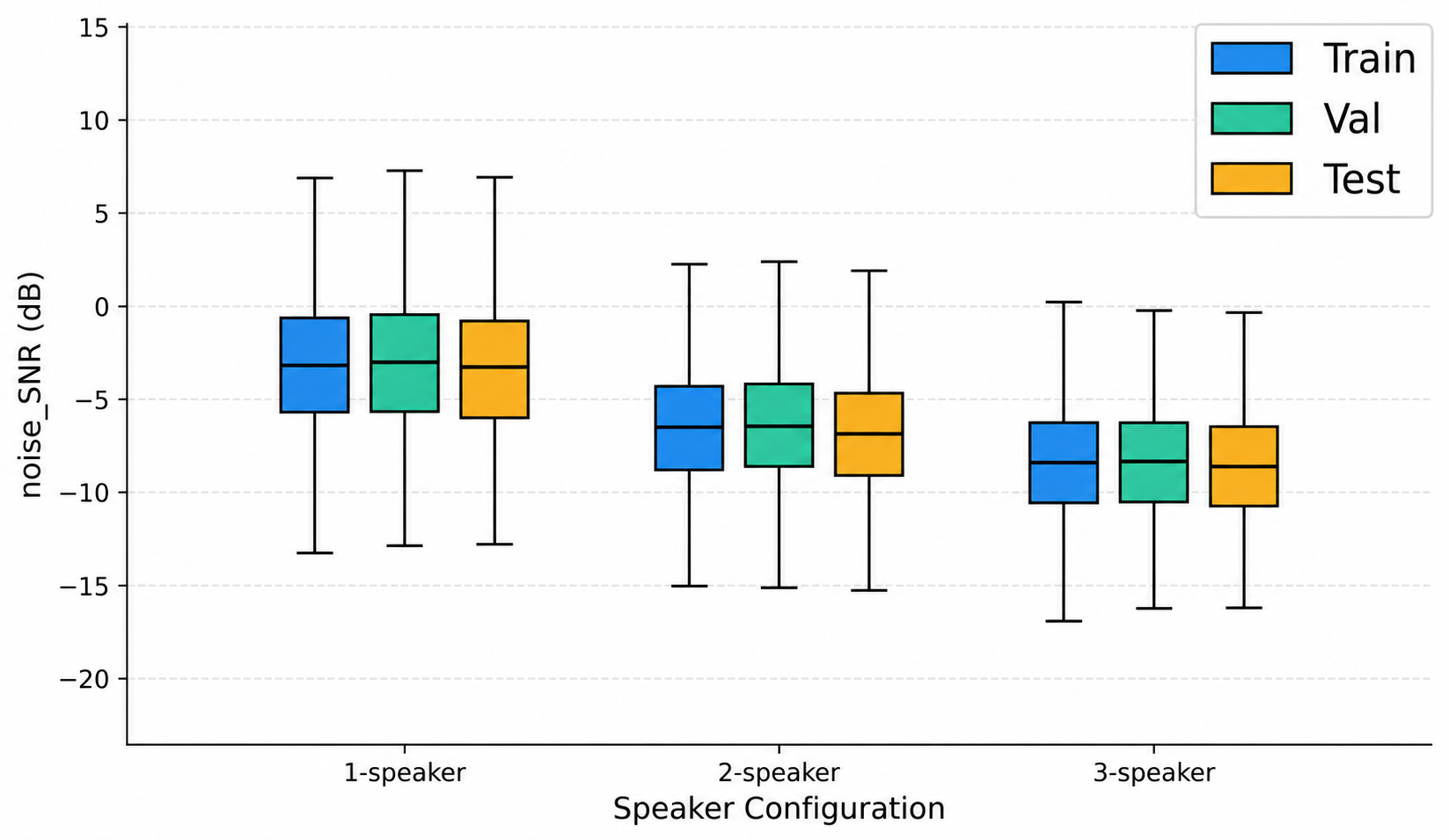}
        \label{fig:boxplot_human}
    }
    \caption{Dataset composition and SNR distributions. (a) Split-wise distribution of original single-speaker recordings and synthesized two-speaker and three-speaker mixtures. (b) SNR distributions under different speaker configurations.}
    \label{fig:data_snr}
\end{figure}

Fig.~\ref{fig:data_bar} further summarizes distributions of original single-speaker recordings and synthesized multi-speaker mixtures. Mixtures are generated independently within each split to preserve speaker disjointness. The distributions of age, gender, and SNR are shown in Fig.~\ref{fig:pies_robot} and Fig.~\ref{fig:boxplot_human}. The SNR diversity arises from stochastic sampling of target loudness levels in Loudness Units relative to Full Scale (LUFS) during LibriMix-style mixture generation, together with inherent energy variation in the source speech and noise recordings.

\section{Methodology}
\begin{figure*}[t]
    \centering
    \includegraphics[width=16cm,height=6cm]{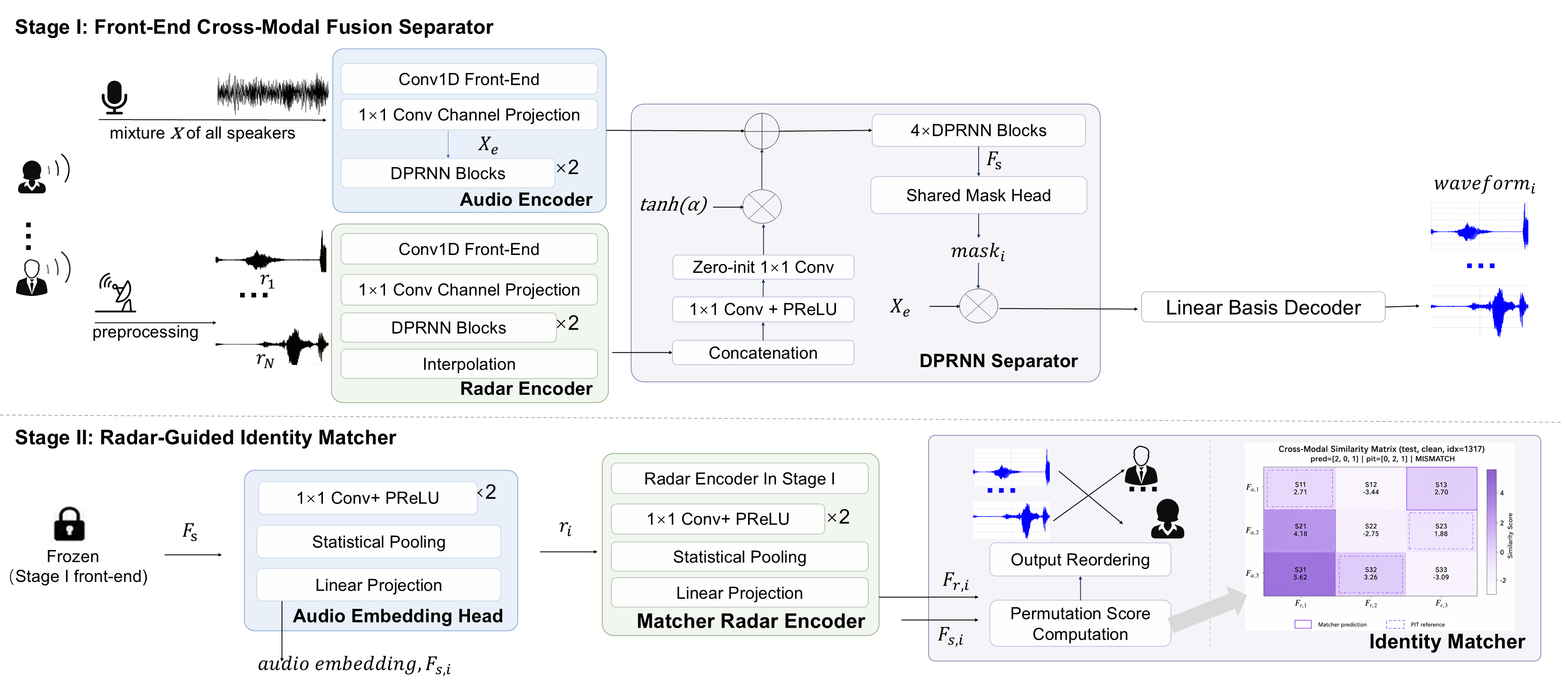}
    \caption{Overall Framework of the Two-Stage Radar-Audio Speech Separation and Cross-Modal Matching Network. Here, $N$ denotes the number of speakers and $i$ indexes the corresponding speaker.}
    \label{fig:model}
\end{figure*}

This section presents the proposed two-stage radar-audio framework for identity-aware speech separation. As shown in  Fig.~\ref{fig:model}, radar information is introduced in two stages: radar-conditioned speech separation and speaker-aware cross-modal matching. Given an acoustic mixture and source-level radar displacement sequences, the framework estimates separated speech streams and assigns each stream to its corresponding radar-observed source.

\subsection{Human-Speech Displacement Refinement}
%\subsubsection{Preprocessing}
Compared with the loudspeaker preprocessing pipeline illustrated in Fig.~\ref{fig:pre} , the human-speech pipeline introduces an additional displacement-refinement step tailored to laryngeal surface motion. Specifically, the human-speech pipeline follows the displacement-extraction procedure in Section III, including chirp-time unfolding, range localization, phase-aligned coherent accumulation, phase unwrapping, and displacement conversion, followed by the newly introduced large-displacement suppression and interpolation step before band-pass filtering. The large-displacement removal is performed on the recovered displacement sequence \(d(m)\) in Eq.~\eqref{eq:displacement}, and is formulated in Eq.~\eqref{eq:clean}. The threshold $\tau$ is set to 200~\(\mu\mathrm{m}\) based on empirical validation and reported micrometer-scale neck-surface displacement \cite{lee2019ultrathin}. This refinement suppresses large non-phonatory motion artifacts.

\begin{equation}
r(t)=
\begin{cases}
d(t), & \left|d(t)-\frac{1}{W}\sum_{i\in\mathcal N_t} d(i)\right|\le\tau,\\[6pt]
\operatorname{Interp}\!\big(d(t)\big), & \text{otherwise},
\end{cases}
\label{eq:clean}
\end{equation}
where \(\mathcal N_t\) denotes the local neighborhood centered at slow-time index \(t\), \(W\) is the baseline window length, \(\tau\) is the displacement threshold, and \(\operatorname{Interp}(\cdot)\) denotes interpolation over the corrupted segments.

\subsection{Radar-Conditioned Speech Separation}
%\subsubsection{Encoders}
\subsubsection{Audio encoder}
The input speech mixture \(x\in\mathbb{R}^{B\times T}\) is first mapped into a low-level acoustic representation \(X_{w}\in\mathbb{R}^{B\times 64\times L}\) through a one-dimensional convolutional front-end. This representation is then projected into a 256-channel separation latent space by a pointwise convolution, yielding \(X_{e}\in\mathbb{R}^{B\times 256\times L}\). The latent sequence \(X_{e}\) is processed by two stacked DPRNN blocks for temporal modeling. Each block alternates intra-segment and inter-segment recurrent modeling on \(256\)-dimensional features. Residual connections and normalization are applied after recurrent modeling for stable optimization. The audio encoder outputs the acoustic representation \(F_{a}\in\mathbb{R}^{B\times 256\times L}\).

\subsubsection{Radar encoder}
Given the radar displacement sequence \(r_i\in\mathbb{R}^{B\times 1\times T_r}\) from the \(i\)-th source, the radar encoder extracts local motion features through a two-layer one-dimensional convolutional front-end. The resulting features are projected into a \(64\)-channel latent space and processed by two stacked DPRNN blocks, yielding a radar representation \(F_{r,i}\in\mathbb{R}^{B\times 64\times L_r}\). Finally, \(F_{r,i}\) is temporally interpolated to align with the acoustic representation for cross-modal fusion.

\subsubsection{Radar-Conditioned Cross-Modal Fusion}
For the \(N\)-speaker case, \(\{F_{r,i}\}_{i=1}^{N}\) are first concatenated along the channel dimension to form a joint radar representation in \(\mathbb{R}^{B\times 64N\times L}\). A pointwise projection maps the joint radar representation into the same 256-dimensional latent space as the acoustic representation \(F_a\in\mathbb{R}^{B\times 256\times L}\). A zero-initialized pointwise convolution then produces the radar-conditioned residual \(\Delta_r\in\mathbb{R}^{B\times 256\times L}\).The fused representation is obtained by injecting \(\Delta_r\) into the acoustic feature through a gated residual connection,
\begin{equation}
F_{\mathrm{fuse}} = F_a + \tanh(\alpha)\,\Delta_r,
\label{eq:fusion}
\end{equation}
with
\begin{equation}
\Delta_r = \Psi\!\left(\mathrm{Concat}(F_{r,1},\dots,F_{r,N})\right),
\label{eq:radar_delta}
\end{equation}
where \(\Psi(\cdot)\) denotes the radar projection module and \(\alpha\) is a learnable fusion gate. 

\subsubsection{Mask Estimation and Decoding}
Given $F_{\mathrm{fuse}}$, a four-layer DPRNN separator estimates latent separation features \(F_{s,i}\in\mathbb{R}^{B\times 256\times L}\) for each output stream. Each output branch predicts a sigmoid-activated soft mask through a lightweight mask head. The estimated mask is applied element-wise to the latent mixture representation \(X_e\in\mathbb{R}^{B\times 256\times L}\), yielding output-specific masked features. The decoder reconstructs time-domain speech streams from the masked features through linear basis transformation and overlap-and-add synthesis, producing separated outputs \(\hat{s}_i\in\mathbb{R}^{B\times T}\).

\subsection{Speaker-Aware Audio-Radar Cross-Modal Matching}
In this second stage, the trained radar-conditioned separator is frozen and produces output-specific features \(F_{s,i}\in\mathbb{R}^{B\times 256\times L}\). These features are mapped into stream-level audio embeddings through pointwise projections, statistical pooling, and a linear projection. On the radar side, an independent matcher encoder maps the \(i\)-th source-level displacement sequence into a radar embedding. The matcher radar encoder shares the convolutional and DPRNN backbone with the radar encoder, but differs in its output objective. Instead of producing temporally aligned sequential features for fusion, it applies pointwise projections, statistical pooling, and a linear projection to obtain a global radar identity embedding. The audio and radar embeddings are normalized, and their inner products form a cross-modal similarity matrix that measures the correspondence between unordered speech streams and radar-observed sources. During training, the PIT-optimal assignment obtained from the frozen separator with respect to the ground-truth speech signals is used as the pseudo target, and the matcher is optimized with a bidirectional matching loss. This design learns cross-modal stream-source alignment while preserving the separation capability of the frozen separator.

\subsection{Training Objective}

\subsubsection{Stage I Front-end Fusion Separator}
For single-speaker enhancement, he radar-conditioned separator is trained by minimizing the negative scale-invariant SI-SDR:
\begin{equation}
\mathcal{L}=-\mathrm{SI\text{-}SDR}(s_1,\hat{s}_1).
\end{equation}
For multi-speaker separation ($K\in\{2,3\}$), permutation-invariant training (PIT) is employed:
\begin{equation}
\mathcal{L}_{\mathrm{PIT}} = 
\min_{\pi\in\Pi_K}\frac{1}{K}\sum_{k=1}^{K}
\left(-\mathrm{SI\text{-}SDR}(s_k,\hat{s}_{\pi(k)})\right),
\end{equation}
where $\Pi_K$ denotes all permutations of $K$ sources.

\subsubsection{Stage II Speaker-Aware Cross-Modal Matcher}

The matcher is trained with a bidirectional matching loss defined over the cross-modal similarity matrix. Let \(S\in\mathbb{R}^{N\times N}\) denote the similarity matrix, where \(S_{ij}\) represents the matching score between the \(i\)-th speech slot and the \(j\)-th radar identity. Let \(\pi\) denote the target permutation inferred from the pseudo labels, and \(\pi^{-1}\) its inverse. The matching loss encourages one-to-one cross-modal correspondence between unordered speech streams and radar identities, and is formulated as
\begin{equation}
\begin{aligned}
\mathcal{L}_{\mathrm{match}}
=
\frac{1}{2N}
\Bigg[
&\sum_{i=1}^{N}
-\log
\frac{\exp\!\big(S_{i,\pi(i)}\big)}
{\sum_{j=1}^{N}\exp\!\big(S_{i,j}\big)}
\\
&+
\sum_{j=1}^{N}
-\log
\frac{\exp\!\big(S_{\pi^{-1}(j),j}\big)}
{\sum_{i=1}^{N}\exp\!\big(S_{i,j}\big)}
\Bigg].
\end{aligned}
\label{eq:matching_loss}
\end{equation}
%The first term enforces each speech slot to be classified to its corresponding radar identity, while the second term imposes the inverse constraint from the radar side to the speech side. 

\section{Experiments}
\subsection{Evaluation Protocol and Metrics}
Separation performance is evaluated with four standard metrics: SI-SDR for waveform reconstruction fidelity, SIR for interference suppression, STOI for speech intelligibility, and PESQ for perceptual speech quality \cite{ozturk2023radio}. For multi-speaker separation, both PIT and ordered evaluations are reported. PIT evaluation measures separation quality under permutation ambiguity, whereas ordered evaluation requires each output stream to match its corresponding radar-observed source. Matching accuracy is used to quantify output-source assignment. For real-world recordings without clean references, WER and CER are reported as reference-free feasibility indicators.

\subsection{Single-Speaker Speech Recovery and Enhancement}
Single-speaker speech recovery or enhancement performance under different sensing modalities and fusion settings are evaluated and presented in Table~\ref{tab:enhance_unseen}. Depending on the available sensing modality, two speech-oriented tasks are considered. 
(i) \textbf{Radar-only} setting is treated as a \emph{speech recovery} task, where speech is reconstructed solely from radar-derived displacement without acoustic input.
(ii) \textbf{Audio-only} and \textbf{Audio-Radar} settings are treated as \emph{speech enhancement} tasks, where noisy input audio is enhanced with or without radar-derived motion cues. 

\begin{table}[t]
\centering
\caption{Results for single-speaker speech recovery and enhancement}
\label{tab:enhance_unseen}
\setlength{\tabcolsep}{6pt}
\renewcommand{\arraystretch}{1.08}
\resizebox{0.9\linewidth}{!}{
\begin{tabular}{l l c c c}
\toprule
\textbf{Modality} & \textbf{Method} &
\textbf{SI-SDR} $\uparrow$ &
\textbf{STOI} $\uparrow$ &
\textbf{PESQ} $\uparrow$ \\
\midrule

\multicolumn{2}{c}{Input (audio)} 
& 3.36 & 0.5983 & 2.045 \\
\multicolumn{2}{c}{Input (radio)} 
& -47.90 & 0.0968 & 1.206 \\
\midrule

\multirow{2}{*}{Audio-Only}
& Conv-TasNet
& 11.79 & 0.723 & \textbf{2.841} \\
& DPRNN
& 12.05 & 0.726 & 2.776 \\
\addlinespace[4pt]

\multirow{2}{*}{Radar-Only}
& WaveEar
& -38.56  & 0.125 & 1.365 \\
& RadioMic
& -50.55 & 0.037 & 1.371 \\
\addlinespace[4pt]

\multirow{3}{*}{Audio-Radar}
& Wavoice
& -40.16 & 0.293 & 1.103 \\
& mmMUSE
& 11.22 & 0.709 & 2.819 \\
& RadioSES
& \textbf{12.07} & \textbf{0.727} & 2.822 \\
& RadarVox
& 11.89 & \textbf{0.727} & 2.786 \\

\bottomrule
\end{tabular}}
\end{table}

\subsubsection{Audio-only baselines}
Compared with the noisy input, Conv-TasNet and DPRNN substantially improve intelligibility and perceptual quality. This is expected because, in the single-speaker setting, the noisy input remains a direct acoustic observation of the target speech. Audio-only models can therefore exploit the acoustic structure of the input and achieve strong enhancement performance.

\subsubsection{Radio-only baselines}
Under the radar-only setting, WaveEar improves SI-SDR by 9.34 dB over the raw radar input, whereas RadioMic underperforms the input. Nonetheless, both methods remain severely limited in SI-SDR and STOI, far from intelligible or high-fidelity speech recovery. This is because radar-measured laryngeal surface motion captures only a partial mechanical correlate of speech rather than the radiated acoustic waveform after vocal-tract shaping. Thus, radar-only sensing is more suitable as an auxiliary cue than as a standalone solution for high-quality speech reconstruction.

\begin{table*}[t]
\centering
\caption{Speech separation results for two-speaker and three-speaker mixtures. Values before and after the slash denote PIT and ordered evaluations, respectively.}
\label{tab:separation_multi}
\setlength{\tabcolsep}{5pt}
\renewcommand{\arraystretch}{1.10}
\begin{tabular}{l cccc cccc}
\toprule
& \multicolumn{4}{c}{\textbf{2-person mix (clean)}} 
& \multicolumn{4}{c}{\textbf{2-person mix (noisy)}} \\
\cmidrule(lr){2-5} \cmidrule(lr){6-9}
\textbf{Model} 
& \textbf{SI-SDR} $\uparrow$ & \textbf{SIR} $\uparrow$ & \textbf{STOI} $\uparrow$ & \textbf{PESQ} $\uparrow$
& \textbf{SI-SDR} $\uparrow$ & \textbf{SIR} $\uparrow$ & \textbf{STOI} $\uparrow$ & \textbf{PESQ} $\uparrow$ \\
\midrule
\multicolumn{1}{l}{Input}
& 0.00 & 0.00 & 0.585 & 1.859
& -1.90 & 1.90 & 0.470 & 1.612 \\
Conv-TasNet
& 8.72 / -5.98 & 20.90 / 6.20 & 0.705 / 0.735 & 2.507 / 2.540
& 5.99 / -7.94 & 22.08 / 8.16 & 0.608 / 0.629 & 2.038 / 2.010 \\
DPRNN
& 9.16 / -6.44 & \textbf{22.13} / 6.52 & 0.718 / \textbf{0.748} & 2.542 / \textbf{2.611}
& 6.28 / -8.36 & 23.09 / 8.46 & 0.622 / \textbf{0.644} & 2.026 / \textbf{2.043}\\
RADIOSES
& 8.75 / 0.60 & 20.54 / 1.31 & 0.707 / 0.586 & 2.478 / 1.890
& 5.99 / -0.46 & 21.49 / 2.18 & 0.614 / 0.501 & 2.021 / 1.671\\
RadarVox
&\textbf{9.75} / \textbf{7.11} & 21.69 / \textbf{18.98} & \textbf{0.728} / 0.675 & \textbf{2.643} / 2.502
&\textbf{6.84} / \textbf{4.51} &  \textbf{23.53} / \textbf{20.46} & \textbf{0.634} / 0.589 & \textbf{2.110} / 1.985\\
\specialrule{1.2pt}{4pt}{4pt}
& \multicolumn{4}{c}{\textbf{3-person mix (clean)}} 
& \multicolumn{4}{c}{\textbf{3-person mix (noisy)}} \\
\cmidrule(lr){2-5} \cmidrule(lr){6-9}
\textbf{Model} 
& \textbf{SI-SDR} $\uparrow$ & \textbf{SIR} $\uparrow$ & \textbf{STOI} $\uparrow$ & \textbf{PESQ} $\uparrow$
& \textbf{SI-SDR} $\uparrow$ & \textbf{SIR} $\uparrow$ & \textbf{STOI} $\uparrow$ & \textbf{PESQ} $\uparrow$ \\
\midrule
\multicolumn{1}{l}{Input}
&-3.35 & -3.34 & 0.451  & 1.547
& -4.35 & -1.40 & 0.395 & 1.096 \\
Conv-TasNet
& 3.10 / -9.94 & 12.03 / 5.62 & 0.584 / 0.339 & 1.821 / 1.469
& 0.60 / -10.09 & 10.97 / 5.53 & 0.501 / 0.293 & 1.655 / 1.406 \\
DPRNN
& 6.85 / -14.69 & 20.85 / 7.32 & 0.655 / 0.227 & 2.278 / 1.440
& 5.16 / -16.24 & 21.97 / 8.12 & 0.600 / 0.213 & 2.014 / 1.403\\
RADIOSES
& 6.65 / -3.24 & 19.61 / -2.94 & 0.654 / 0.450  & 2.309 / 1.538
& 5.06 / -3.75 & 20.76 / -1.94 & 0.599 / 0.411 & 2.053 / 1.495 \\
RadarVox
&\textbf{7.13} / \textbf{5.43} & \textbf{22.15} / \textbf{17.83} & \textbf{0.690} / \textbf{0.630} & \textbf{2.356} / \textbf{2.278}
& \textbf{5.49} / \textbf{3.97} & \textbf{23.92} / \textbf{19.51} & \textbf{0.636} / \textbf{0.582} & \textbf{2.070} / \textbf{2.012}\\
\bottomrule
\end{tabular}
\end{table*}

\subsubsection{Audio and radar fusion methods}
For Wavoice, a Griffin–Lim-based inversion module is applied to its log-mel outputs to recover speech waveforms. Since Wavoice is designed for speech recognition rather than enhancement, this inversion step may introduce spectral inconsistency and reconstruction artifacts, resulting in degraded performance in Table~\ref{tab:enhance_unseen}. In contrast, mmMUSE adopts complex-valued modeling, attention-based cross-modal fusion, and time--frequency masking. RadioSES treats mmWave as a modality parallel to audio and achieves the highest SI-SDR of 12.07 dB. Our method achieves comparable performance, indicating that the proposed framework can exploit radar as a complementary motion cue for speech enhancement.

\subsection{Multi-Speaker Speech Separation}
We evaluate multi-speaker separation from two perspectives: PIT-based separation quality and ordered identity-aware performance. Table~\ref{tab:separation_multi} reports both PIT and ordered metrics under clean and noisy two- and three-speaker settings.

\subsubsection{PIT-Based Separation Quality}
%Under PIT evaluation, i.e., the values reported before the slash in Table~\ref{tab:separation_multi}, the focus is on separation quality. 
The three-speaker setting is more challenging due to stronger interference and increased overlap, but the overall trends remain consistent with the two-speaker case. RadarVox slightly outperforms the strong audio-only DPRNN baseline, while RadioSES shows lower multi-speaker separation performance. The result indicates that radar-derived motion cues provide complementary information for separating overlapped speech.

\subsubsection{Speaker Identity Assignment}
Under ordered evaluation, each separated stream must be aligned with its corresponding radar-observed source. Except for our RadarVox model, other models exhibit sharp degradation from PIT to ordered metrics, indicating that separation quality alone does not ensure reliable output-source correspondence. By leveraging radar-derived motion cues and cross-modal matching, RadarVox model substantially reduces the PIT-to-ordered gap and achieves matching accuracies of approximately 88\% and 83\% for two- and three-speaker mixtures, respectively.

\subsection{Ablation Analysis}
\begin{figure}[htbp]
    \centering
    \includegraphics[width=\linewidth]{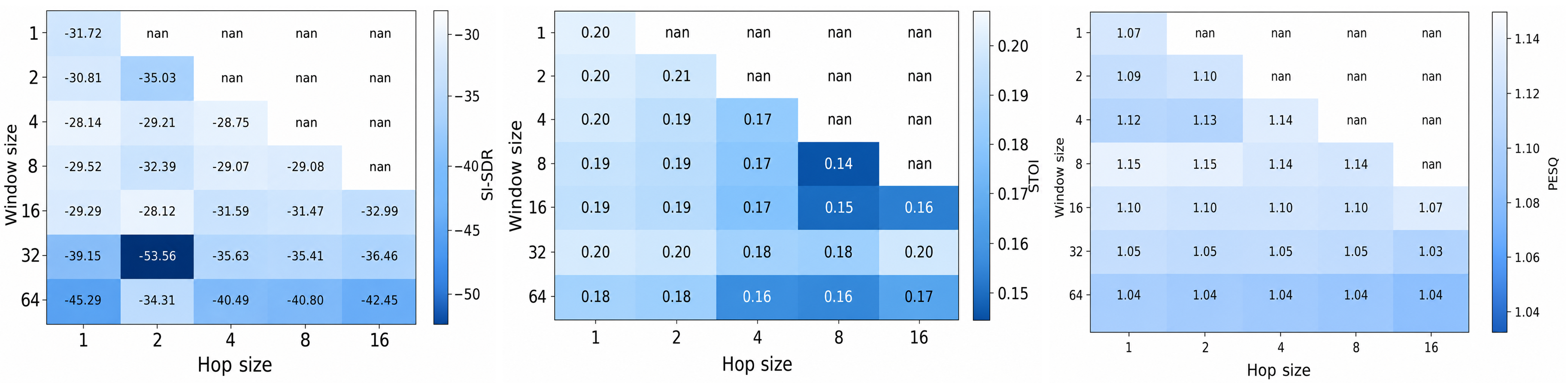}
    \vspace{2pt}
    \makebox[\linewidth][c]{%
      \makebox[0.33\linewidth][c]{\hspace{-0.02\linewidth}(a)}%
      \makebox[0.33\linewidth][c]{(b)}%
      \makebox[0.33\linewidth][c]{(c)}%
   }
    \caption{Speech quality heatmaps under different window and hop sizes. In the SI-SDR, STOI, and PESQ heatmaps, lighter colors consistently indicate better performance.}
    \label{fig:ab_0}
\end{figure}

\begin{figure}[htbp]
    \centering
    \includegraphics[width=0.85\linewidth,height=0.34\linewidth]{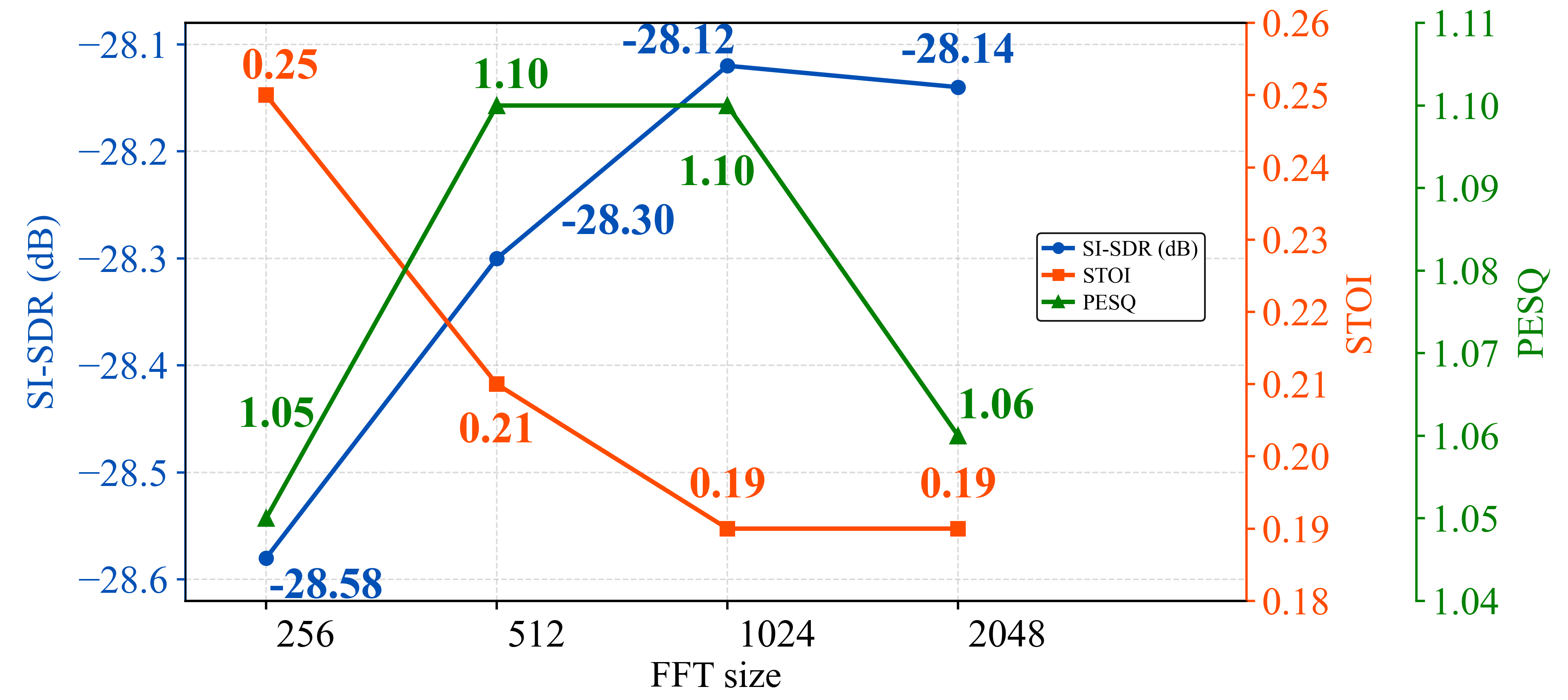}
    \caption{Speech quality evaluation under different FFT size}
    \label{fig:ab_01}
\end{figure}

Ablation analysis evaluates three key components: preprocessing parameters, radar-conditioned fusion, and speaker-aware cross-modal matching.
Preprocessing analysis in Fig.~\ref{fig:ab_0} and Fig.~\ref{fig:ab_01} shows that a window size of \(4\), hop size of \(1\), and FFT size of \(1024\) provide a favorable trade-off between global temporal structure and transient speech cues.

 \begin{table}[htbp]
\centering
\scriptsize
\caption{Ablation of Stage I radar integration under PIT evaluation.}
\label{tab:stage1_pit}
\setlength{\tabcolsep}{2.0pt}
\renewcommand{\arraystretch}{0.92}
\resizebox{0.9\linewidth}{!}{
\begin{tabular}{llcccccccc}
\toprule
\multirow{2}{*}{Spk.} & \multirow{2}{*}{Setting}
& \multicolumn{4}{c}{Clean}
& \multicolumn{4}{c}{Noisy} \\
\cmidrule(lr){3-6} \cmidrule(lr){7-10}
& & SI-SDR & SIR & STOI & PESQ
  & SI-SDR & SIR & STOI & PESQ \\
\midrule
\multirow{2}{*}{2spk}
& w/o radar & 9.16 & \textbf{22.13} & 0.718 & 2.542 & 6.28 & 23.09 & 0.622 & 2.026 \\
& w/ Radar & \textbf{9.75} & 21.69 & \textbf{0.728} & \textbf{2.643} & \textbf{6.84} & \textbf{23.53} & \textbf{0.634} & \textbf{2.110} \\
\cmidrule(lr){1-10}
\multirow{2}{*}{3spk}
& w/o radar & 6.85 & 20.85 & 0.655 & 2.278 & 5.16 & 21.97 & 0.600 & 2.014 \\
& w/ Radar & \textbf{7.13} & \textbf{22.15} & \textbf{0.690} & \textbf{2.356} & \textbf{5.49} & \textbf{23.92} & \textbf{0.636} & \textbf{2.070} \\
\bottomrule
\end{tabular}}
\end{table}

\begin{table}[htbp]
\centering
\scriptsize
\caption{Ablation of Stage I radar integration under ordered evaluation.}
\label{tab:stage1_order}
\setlength{\tabcolsep}{2.0pt}
\renewcommand{\arraystretch}{0.92}
\resizebox{0.9\linewidth}{!}{
\begin{tabular}{llcccccccc}
\toprule
\multirow{2}{*}{Spk.} & \multirow{2}{*}{Setting}
& \multicolumn{4}{c}{Clean}
& \multicolumn{4}{c}{Noisy} \\
\cmidrule(lr){3-6} \cmidrule(lr){7-10}
& & SI-SDR & SIR & STOI & PESQ
  & SI-SDR & SIR & STOI & PESQ \\
\midrule
\multirow{2}{*}{2spk}
& w/o Radar & -6.44 & \textbf{6.52} & \textbf{0.748} & \textbf{2.611} & -8.36 & \textbf{8.46} & \textbf{0.644} & \textbf{2.043} \\
& w/ Radar & \textbf{-0.03} & -0.01 & 0.588 & 1.899 & \textbf{-0.96} & 0.00 & 0.499 & 1.669 \\
\cmidrule(lr){1-10}
\multirow{2}{*}{3spk}
& w/o Radar & \textbf{-14.69} & 7.32 & 0.227 & 1.440 & \textbf{-16.24} & 8.12 & 0.213 & 1.403 \\
& w/ Radar & -15.27 & \textbf{7.36} & \textbf{0.293} & \textbf{1.614} & -17.41 & \textbf{8.39} & \textbf{0.276} & \textbf{1.523} \\
\bottomrule
\end{tabular}}
\end{table}

 As shown in Table~\ref{tab:stage1_pit}, introducing radar consistently improves noise suppression, multi-speaker separation, intelligibility, and perceptual quality. However, Table~\ref{tab:stage1_order} shows that, without the matcher, Stage-I radar fusion provides a weak ordering bias rather than explicit identity assignment. In the two-speaker case, this bias can partially break speaker symmetry and alleviate binary speaker swaps, improving ordered SI-SDR from -6.44 dB to -0.03 dB. In the three-speaker case, however, the assignment problem becomes a full three-way permutation, and the weak ordering cues from Stage I are no longer sufficient to establish stable speaker-output correspondence.

%\begin{figure}[htbp]
    %\centering
    %\includegraphics[width=\linewidth]{ab_1.png}
    %\caption{Stage I radar integration under PIT or Ordered Evaluation}
    %\label{fig:ab_1}
%\end{figure}

For Table~\ref{tab:stage2_ordered}, matching accuracy is evaluated using the frozen Stage I outputs. The permutation yielding the best PIT SI-SDR is treated as the pseudo ground-truth ordering, and the matcher is considered correct when its highest-scoring permutation matches this ordering. The results show that the matcher achieves over 80\% accuracy in both the two-speaker and three-speaker cases, supporting the effectiveness of radar-guided identity-aware ordered separation.

\begin{table}[htbp]
\centering
\scriptsize
\caption{Ablation of Stage II matcher learning under ordered evaluation.}
\label{tab:stage2_ordered}
\setlength{\tabcolsep}{1.6pt}
\renewcommand{\arraystretch}{0.88}
\resizebox{\columnwidth}{!}{
\begin{tabular}{llcccccccccc}
\toprule
\multirow{2}{*}{Spk.} & \multirow{2}{*}{Setting}
& \multicolumn{5}{c}{Clean}
& \multicolumn{5}{c}{Noisy} \\
\cmidrule(lr){3-7} \cmidrule(lr){8-12}
& & \raisebox{0.6ex}{SI-SDR}
    & \raisebox{0.6ex}{SIR}
    & \raisebox{0.6ex}{STOI}
    & \raisebox{0.6ex}{PESQ}
    & \shortstack[c]{Match\\Acc.}
    & \raisebox{0.6ex}{SI-SDR}
    & \raisebox{0.6ex}{SIR}
    & \raisebox{0.6ex}{STOI}
    & \raisebox{0.6ex}{PESQ}
    & \shortstack[c]{Match\\Acc.} \\
\midrule
\multirow{2}{*}{2spk}
& w/o Matcher & -0.03 & -0.01 & 0.588 & 1.899 & 49.9\% & -0.96 & 0.00 & 0.499 & 1.669 & 49.9\%\\
& w/ Matcher  & \textbf{7.11} & \textbf{18.98} & \textbf{0.675} & \textbf{2.502} & \textbf{88.2\%} & \textbf{4.51} & \textbf{20.46} & \textbf{0.589} & \textbf{1.985} & \textbf{88.9\%} \\
\cmidrule(lr){1-12}
\multirow{2}{*}{3spk}
& w/o Matcher & -15.27 & 7.36 & 0.293 & 1.614 & 16.3\% &-17.41 & 8.39 & 0.276 & 1.523 & 16.4\%\\
& w/ Matcher  &\textbf{5.43} & \textbf{17.83} & \textbf{0.630} & \textbf{2.278} & \textbf{82.8\%} & \textbf{3.97} & \textbf{19.51} & \textbf{0.582} & \textbf{2.012} & \textbf{82.5\%} \\
\bottomrule
\end{tabular}}
\end{table}

Additional real-world case studies and radar-only emotion recognition experiments are provided in the supplementary material to further examine affective sensing potential and extended benchmark utility.

\section{Conclusion}
In this work, we address identity-aware cocktail-party speech separation under radar-audio sensing and establish RadarVox, a benchmark consisting of a publicly available radar-audio dataset and a two-stage framework for this task. RadarVox also includes a preprocessing pipeline for extracting speech-related micro-vibration cues from raw FMCW measurements. The proposed framework integrates radar-gated fusion for robust separation and cross-modal matching for explicit speaker assignment. Experimental results show consistent gains in multi-speaker separation quality and substantial improvements in ordered identity-aware evaluation. These findings suggest that radar-derived motion cues can serve as complementary physical information for identity-aware speech perception in embodied voice-interaction environments.

\bibliographystyle{IEEEtran}
\bibliography{IEEEabrv,references}

\end{document}